\documentclass[12pt]{iopart}

\usepackage{graphicx}
\begin{document}

\title{A novel pulsed fiber laser: Further study on the bias-pumped gain-switched fiber laser}

\author{Fuyong Wang}

\address{School of Information and Electrical Engineering, Hebei University of Engineering, Handan 056038, China}
\ead{jiaoyi@sjtu.edu.cn}
\vspace{10pt}
\begin{indented}
\item[]April 2018
\end{indented}

\begin{abstract}
The bias-pumped gain-switched fiber laser proposed by us is considered a novel pulsed fiber laser based on a new pulsing mechanism. With a certain signal power seeding, synchronization of temporal evolution can be kept between the output signal laser and the pump. The seed laser can be supplied conveniently by a CW pump power which is named bias pump power. A pulsed pump is responsible for shaping the output pulse. Stable pulsed lasers with tunable durations can be achieved under bias pump combined with pulsed pump. In addition, the temporal shape of output pulses can be controllable based on this new pulsing mechanism. Compared with conventional gain-switched fiber laser, a much simpler pulsed laser design can be provided by this novel pulsed fiber laser because it is no need to add a control unit to realize fast gain-switching.
\end{abstract}

%
%
%
%
%

\section{Introduction}

Sustained interest in the pulsed lasers is motivated by their widespread use in micromachining applications. Q-switching and gain-switching are the two common approaches to achieve pulsed operation realizing laser pulse with duration from tens of nanoseconds to several microseconds \cite{Maryashin2006,Swiderski2004,Larsen2014}. Compared with Q-switching, gain-switching method offers an more compact and robust way of building a laser system. There is no additional active optical elements required for modulation inside the cavity. Relying on fast modulation of the pump power, gain-switching of a laser is considered as the most robust and simplest way to produce nanosecond laser pulses \cite{Fengqiu2016}. Combined with fiber technology, gain-switched fiber laser provides a compact and simple setup, especially in all-fiber construction \cite{Vid2014,intro1,intro2,introt,one1,one2,one4,one5,intro3}.

The temporal characteristics of gain-switched pulse including duration and peak power are strongly related to the time-varying of pump power \cite{Fengqiu2016}. Obtaining a stable pulse that is the first spike of relaxation oscillation requires switching off the pump power at an appropriate time. That is why fast gain-switching techquine is generally adopted in gain-switched fiber laser to suppress the following spikes after the initial one \cite{one3,Swiderski2013}. Although it is no need to add active optical elements required for modulation inside the cavity, a much more complex feedback loop is required to realize fast gain-switching. 

In recent years pulsed lasers with tunable durations are highly favored in various applications \cite{Sikora2010,Kwang2012,Pascal2009}. To satisfy different applications, it typically requires pulsed lasers with different durations and energies. Besides, the temporal shapes of pulsed lasers are also of importance in laser applications \cite{Pascal2009}. For instance, in the case of laser micromachining, the material removal efficiency depends on the temporal shapes and durations of pulsed lasers \cite{Pascal2009}. Consequently, pulse-duration-tunable and temporal-profile-controllable lasers are essential in order to ensure flexible and precise processing. However, the duration of the stable pulse output from gain-switched fiber laser is limited by that of pump pulse because chaotic relaxation spikes will appear when the pump duration is long, which in some cases limits the applications of gain-switched fiber laser. What is worse, the output pulse shape is hard to be controlled in present gain-switched fiber lasers.

In a previous work, we demonstrated both long pump duration and high peak power can result in chaotic relaxation spike phenomenon, and proposed a bias pumping technique to regulate the spikes in gain-switched fiber laser \cite{Fuyong2018}. However, the mechanism of regulating chaotic spikes by a certain pump power bias is not explained. In addition, some new effects in the bias-pumped gain-switched fiber laser have not been investigated detailly. 

In this letter, we try to explain why chaotic spikes can be suppressed with a certain pump power bias and reveal the pulsing mechanism of this novel fiber laser is completely different with that of conventional gain-switched fiber laser. Based on a new pulsing mechanism, we explore whether stable output pulses with tunable durations can be generated and the temporal profile of output pulse can be shaped as required. We also show this novel pulsed fiber laser can provide a much simpler pulsed laser design than conventional gain-switched fiber laser, since the feedback loop in conventional gain-switched fiber laser used to achieve fast gain-switching is not necessary in this novel pulsed fiber laser. 

\section{Numerical analysis of relaxation oscillation based on traveling-wave rate equations}

Pulsed outputs of gain-switched fiber are due to relaxation oscillation. Stable gain-switched pulse is the first isolated spike of the relaxation oscillation. The key factor for obtaining stable pulse is to design the pump source to achieve fast gain-switching which can suppress the following spikes after the initial one. To explore method of controlling the temporal properties of output pulses in gain-switched fiber laser, it is necessary to analyse the temporal characteristics of relaxation oscillation firstly. Choosing Yb-doped fiber as a gain medium, relaxation oscillation is simulated based on traveling-wave rate equations given by \cite{two1}
\begin{equation}
N=N_1+N_2,\label{e1}
\end{equation}
\begin{equation}
\frac{\partial N_2}{\partial t}+\frac{N_2}{\tau }=\frac{\Gamma _p\lambda _p}{hcA}[\sigma _{ap}N_1-\sigma _{ep}N_2]P_p+\frac{\Gamma _s\lambda _s}{hcA}[\sigma _{as}N_1-\sigma _{es}N_2](P_{sf}+P_{sb}),\label{e2}
\end{equation}
\begin{equation}
\frac{\partial P_p}{\partial z}+\frac{1}{\upsilon _p}\frac{\partial P_p}{\partial t}=\Gamma _p[\sigma _{ep}N_2-\sigma _{ap}N_1]P_p-\alpha _pP_p,
\end{equation}
\begin{equation}
\frac{\partial P_{sf}}{\partial z}+\frac{1}{\upsilon _s}\frac{\partial P_{sf}}{\partial t}=\Gamma _s[\sigma _{es}N_2-\sigma _{as}N_1]P_{sf}-\alpha _sP_{sf}+2\sigma _{es}N_2\frac{hc^2}{\lambda _{s}^3}\Delta \lambda _s, 
\end{equation}
\begin{equation}
-\frac{\partial P_{sb}}{\partial z}+\frac{1}{\upsilon _s}\frac{\partial P_{sb}}{\partial t}=\Gamma _s[\sigma _{es}N_2-\sigma _{as}N_1]P_{sb}-\alpha _sP_{sb}+2\sigma _{es}N_2\frac{hc^2}{\lambda _{s}^3}\Delta \lambda _s,\label{e5}
\end{equation}
where N$_1$, N$_2$ and N are the lower and upper population concentrations and the total doping concentration, respectively. h, $\tau$ and c are the Planck constant, fluorescence lifetime and the speed of light in a vacuum, respectively. $\Gamma _s$ and $\lambda _s$ are the overlap factor between the signal and the doped fiber area and the signal free-space wavelength, respectively. $\Gamma _p$ and $\lambda _p$ are the overlap factor between the pump and the doped fiber area and the pump free-space wavelength, respectively.
A is the core area of the fiber. $\sigma _{es}$ and $\sigma _{as}$ are the emission and absorption cross sections of the signal power, respectively. $\sigma _{ep}$ and $\sigma _{ap}$ are the emission and absorption cross sections of the pump power, respectively. The attenuation of the signal and pump powers are represented by $\alpha _s$ and $\alpha _p$, respectively. P$_{sb}$, P$_{sf}$ and P$_p$ are the backward, forward signal powers and pump power, respectively. $\upsilon _s$ and $\upsilon _p$ are group velocities of the signal laser and pump propagating in the fiber, respectively. $\Delta \lambda _s$ is the bandwidth of the amplified spontaneous emission (ASE).

The rate equations (\ref{e1})-(\ref{e5}) are governed by the boundary conditions physically representing the feedback provided by the reflectors on either end. The typical boundary conditions for equations (\ref{e1})-(\ref{e5}) are given by
\begin{equation}
P_p(z=0,t)=W_0,\label{eb1}
\end{equation}
\begin{equation}
P_{sf}(z=0,t)=R_1P_{sb}(z=0,t),\label{eb2}
\end{equation}
\begin{equation}
P_{sb}(z=L,t)=R_2P_{sf}(z=L,t),\label{eb3}
\end{equation}
\begin{equation}
P_{out}(z=L,t)=(1-R_2)P_{sf}(z=L,t),\label{eb4}
\end{equation}
where R$_1$ and R$_2$ are the reflectivities of reflector 1 and reflector 2, respectively, and P$_{out}$ is the output signal power. Details of the parameters used in the simulation are summarized in Table \ref{t1}. 
\begin{table}[htbp]
\newcommand{\tabincell}[2]{\begin{tabular}{@{}#1@{}}#2\end{tabular}}
  \caption{\label{t1}Detail parameters used in the simulations \cite{two1}}
  \begin{center}
    \begin{tabular}{ccccccccc}
    \hline
     \tabincell{c}{Parameter}& Value& Parameter& Value  \\
    \hline
    $\lambda _p$ &980 nm &$\sigma _{ap}$ &2.5$\times 10^{-24}$ m$^2$ \\
    $\lambda _s$ & 1090 nm&$\alpha _s$ & 5$\times 10^{-3}$ m$^{-1}$ \\
    N & 2$\times 10^{26}$ m$^{-3}$&$\alpha _p$ & 0.39 m$^{-1}$ \\
    $\tau$  & 1 ms&$\Delta \lambda _s$ & 20 nm\\
    A &2.83$\times 10^{-11}$ m$^2$& $\Gamma _s$  & 0.75\\
    $\sigma _{es}$ & 3.5$\times 10^{-25}$ m$^2$&$\Gamma _p$ & 0.0023\\
    $\sigma _{as}$ &2.0$\times 10^{-27}$ m$^2$& $R_1$ & 0.99\\
    $\sigma _{ep}$ & 3.0$\times 10^{-24}$ $m^2$&$R_2$  & 0.1 \\
    \hline
    \end{tabular}
  \end{center}
\end{table}

\begin{figure*}[htbp]
\centerline{
\includegraphics[width=8cm]{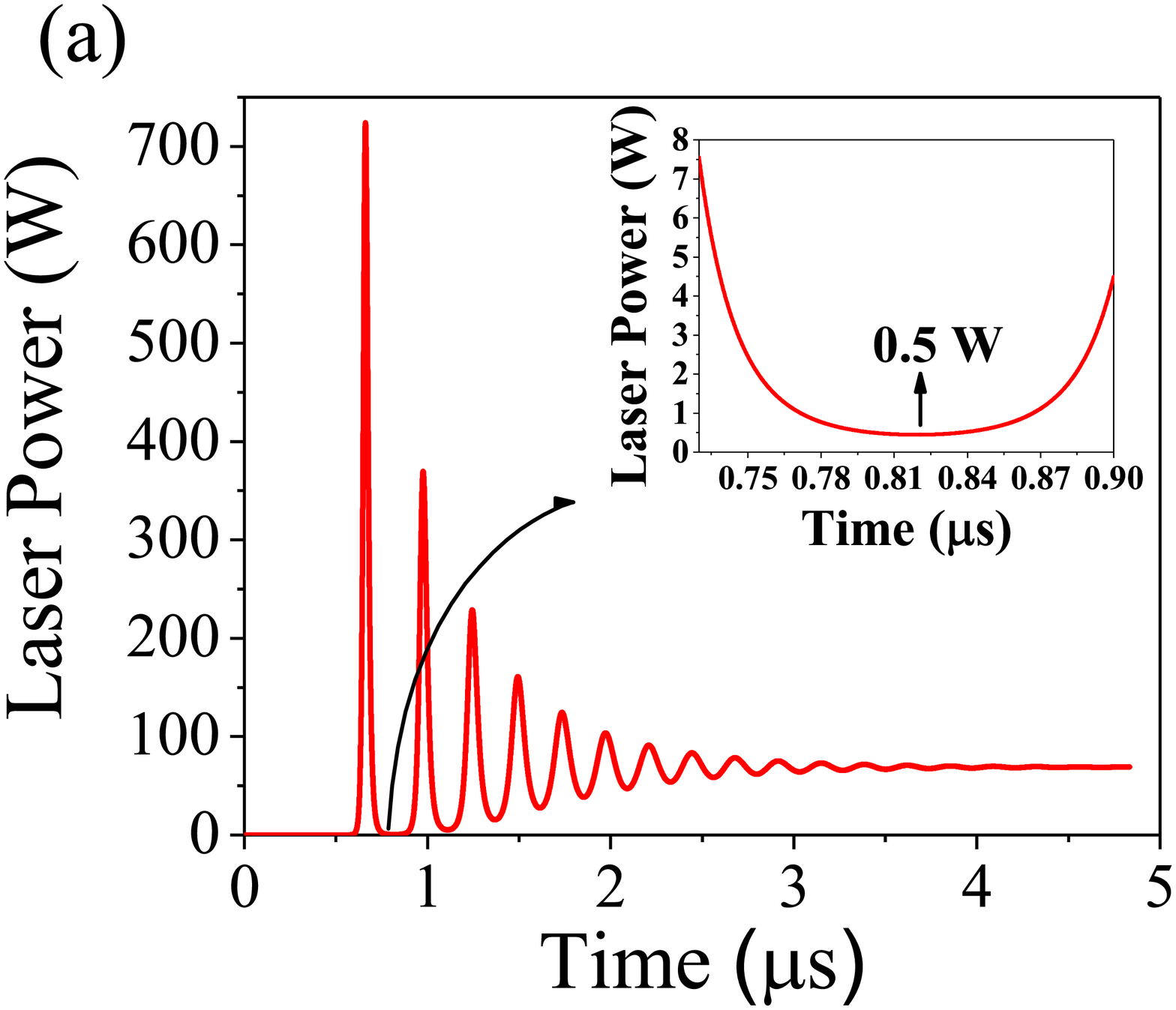}
\includegraphics[width=8cm]{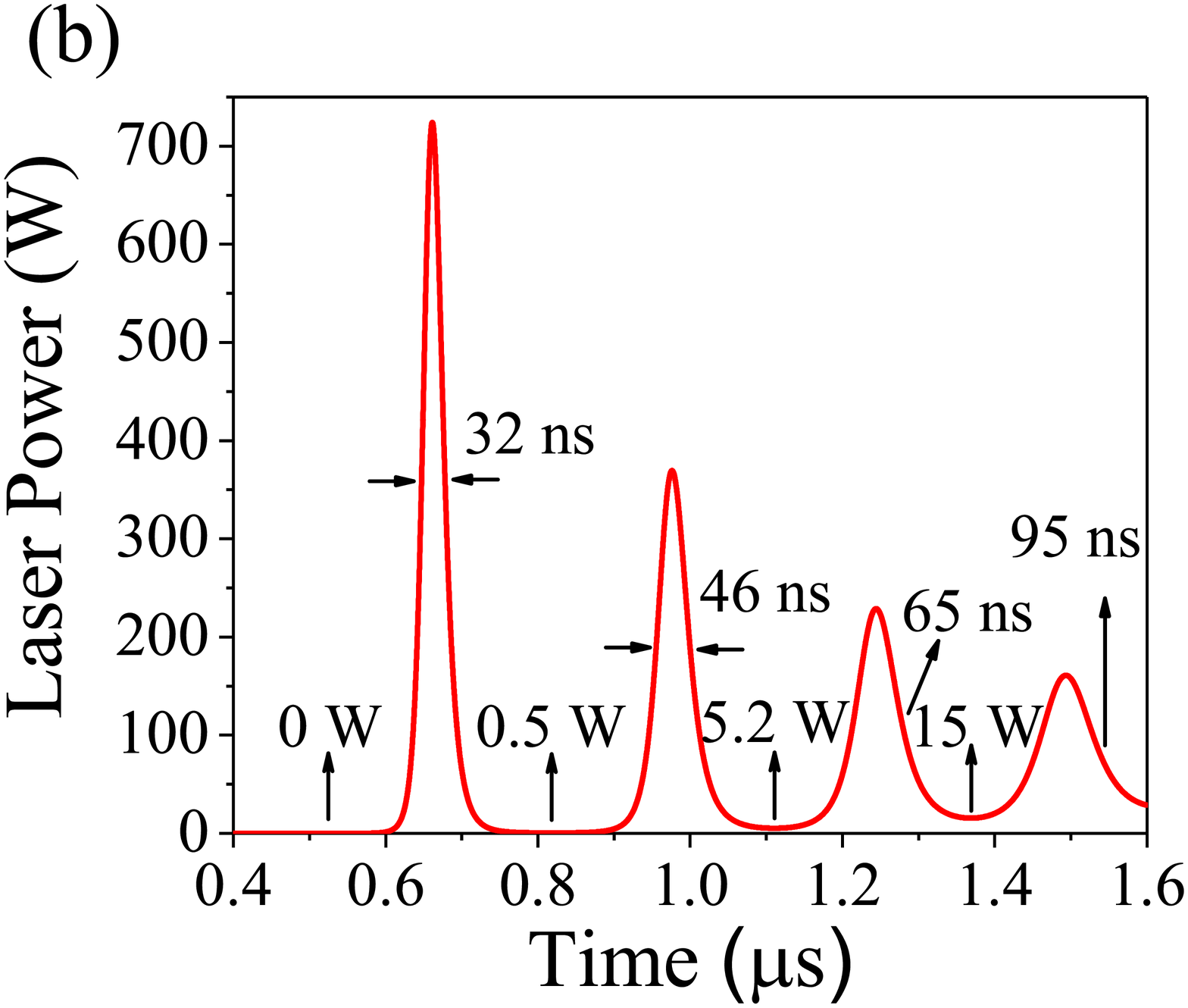}}
\caption{Under CW pumping power of 100 W: (a) temporal characteristics of output lasers, (b) the first four spikes of relaxation oscillation.}\label{pf}
\end{figure*}

Figure \ref{pf} is the simulation result based on rate equations (\ref{e1})-(\ref{e5}) and the boundary conditions (equations (\ref{eb1})-(\ref{eb4})). When the continuous-wave (CW) pumping power is above the threshold, relaxation oscillation will inevitablly occur before the output laser reaches a steady state, as seen in Fig. \ref{pf}(a). The relaxation oscillation consists of several spikes. With power starting from zero the first spike has the highest peak power and the narrowest duration. Seen from the inset of Fig. \ref{pf}(a), the laser power after the first spike doesnot decrease to zero and then the second spike appears. It means that the second spike is seeded with a certain amount of laser power (about 0.5 W). The seeded power (about 5.2 W) of the third spike is higher than that of the second one, as shown in Fig. \ref{pf}(b). The laser power at the interval of two adjacent spikes, which severs as a seed for the later one, becomes higher and higher with time. Seeded with a higher power, the duration (amplitude) of the spike is wider (lower) than those of the previous one. Therefore, it is natural to ask whether it is the increase of the seed power that results in the decrement of the amplitude and broadening of the duration of the following spike. In the following, we will study on the effects caused by the seeded laser based on the parameters in Table \ref{t1}.

\begin{figure*}[htbp]
\centerline{\includegraphics[width=4in]{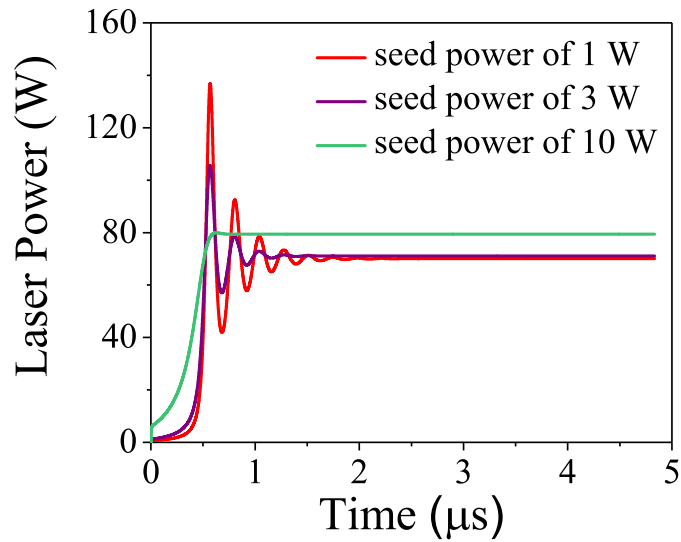}}
\caption{Under different seeded powers temporal characteristics of output lasers in the seeded fiber laser with CW pump power of 100 W.}\label{ps}
\end{figure*}

With a certain power (W$_s$) of signal laser continuously seeding into the fiber from the same side as the pump injected (at z=0), the boundary condition of the forward signal power is changed to $P_{sf}(z=0,t)=R_1P_{sb}(z=0,t)+W_s$. The simulation results of the seeded fiber laser are shown in Fig. \ref{ps}.

Seeded with different signal powers, the temporal characteristics of outputs from CW pumped fiber laser are depicted in Fig. \ref{ps}. Compared with Fig. \ref{pf}(a) the amplitude of relaxation oscillation spikes are greatly reduced with signal laser seeding, as shown in Fig. \ref{ps}. Both the amplitudes and the number of relaxation spikes decrease gradually with the increase of the seeded power. That is to say, the higher the seeded power, the shorter the lasting time of relaxation oscillation becomes, and the earlier the CW steady state of output is to be reached. Consequently, we have reasons to believe that the relaxation oscillation can be completely eliminated when the seeded power increases to a certain value. If the fiber laser is seeded with signal power of 10 W continuously, the output laser quickly (after 0.5 $\mu$s) reaches a CW state without experiencing relaxation oscillation, as shown in the green line of Fig. \ref{ps}. Therefore, we can safely conclude that the seeded power of signal can mitigate the relaxation oscillation of output laser. 

In our previous work, we have demonstrated the chaotic spikes of gain-switched fiber laser can be eliminated with bias-pumping technique \cite{Fuyong2018}. Actually, the bias-pumped power is responsible for generating signal laser at the interval of two adjacent pulses. It is the CW signal laser, severing as a seed to the next pulse, produced by bias-pump power, that in reality regulates the chaotic spikes of outputs in the bias-pumped gain-switched fiber laser. Thus, adpoting bias-pumping is a simple approach to implement laser seeding. Obviously, the pulsing mechanism of the bias-pumped gain-switched fiber laser is not based on relaxation oscillation. In this sense, it is a novel pulsed fiber laser based on a new pulsing mechanism, which deserves to be further investigated.

\section{Pulse-shaping in the bias-pumped gain-switched fiber laser}

\begin{figure*}[htbp]
\centerline{\includegraphics[width=5in]{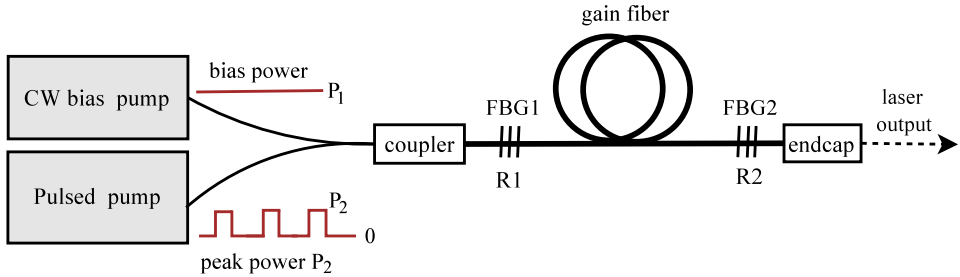}}
\caption{Schematic representation of the bias-pumped gain-switched fiber laser.}\label{setup}
\end{figure*}

Figure \ref{setup} is a schematic of the bias-pumped gain-switched fiber laser. The linear cavity consists of Yb-doped double-cladding fiber with a length of 1 m and fiber bragg grating (FBG) pairs. Pulsed pump and CW bias pump constitute the pump source of the bias-pumped gain-switched fiber laser. Both the pulsed and CW bias pump source consist of several high-power laser diodes which are readily available for the Yb-doped fiber. Using the main parameters in Table. \ref{t1}, we simulate how stable pulses are produced in this novel fiber laser. 

\begin{figure*}[htbp]
\begin{minipage}{0.5\linewidth}
\centerline{\includegraphics[width=7cm]{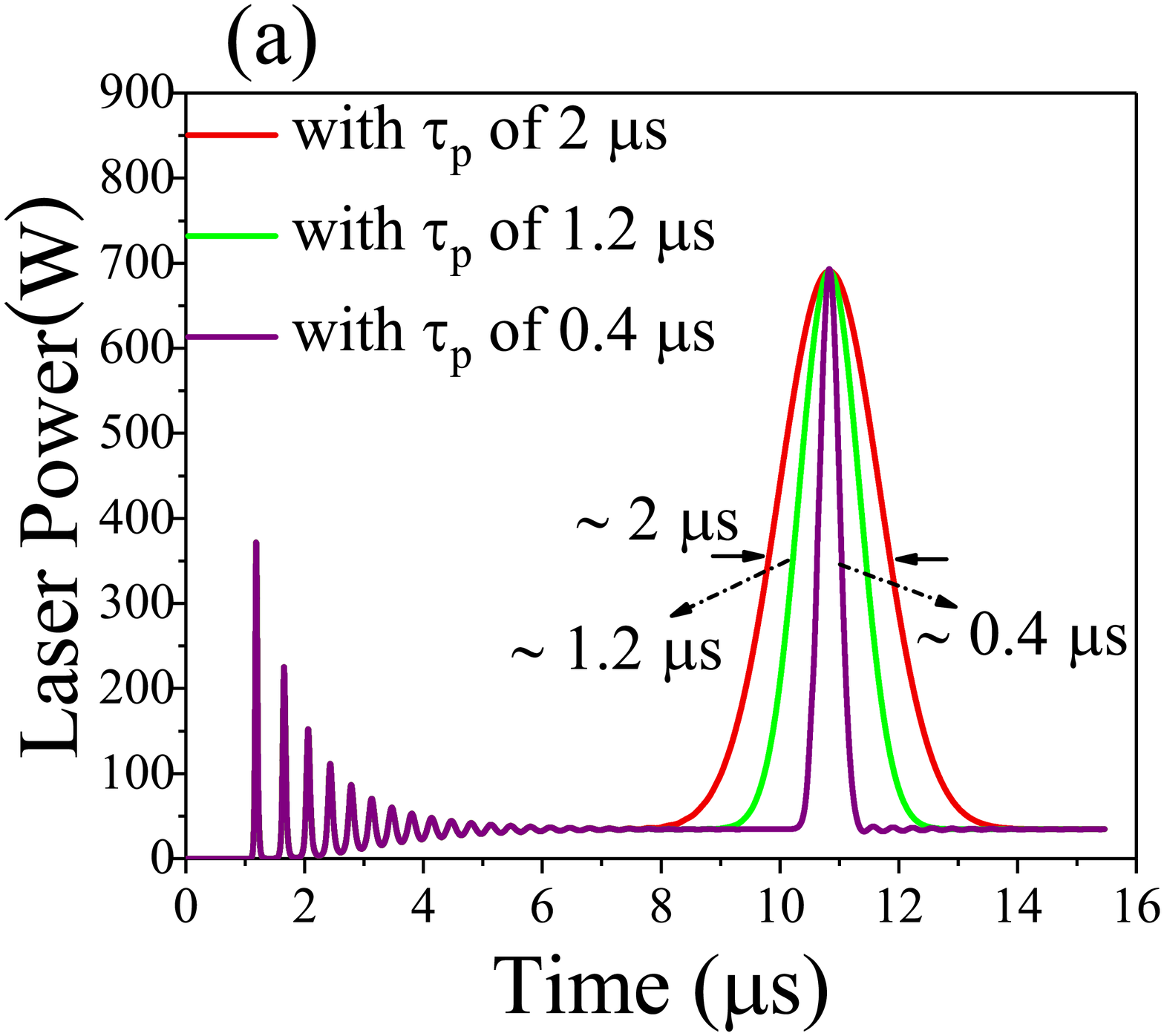}}
\centerline{ }
\end{minipage}
\hspace{-0.35in}
\begin{minipage}{0.5\linewidth}
\centerline{\includegraphics[width=7cm]{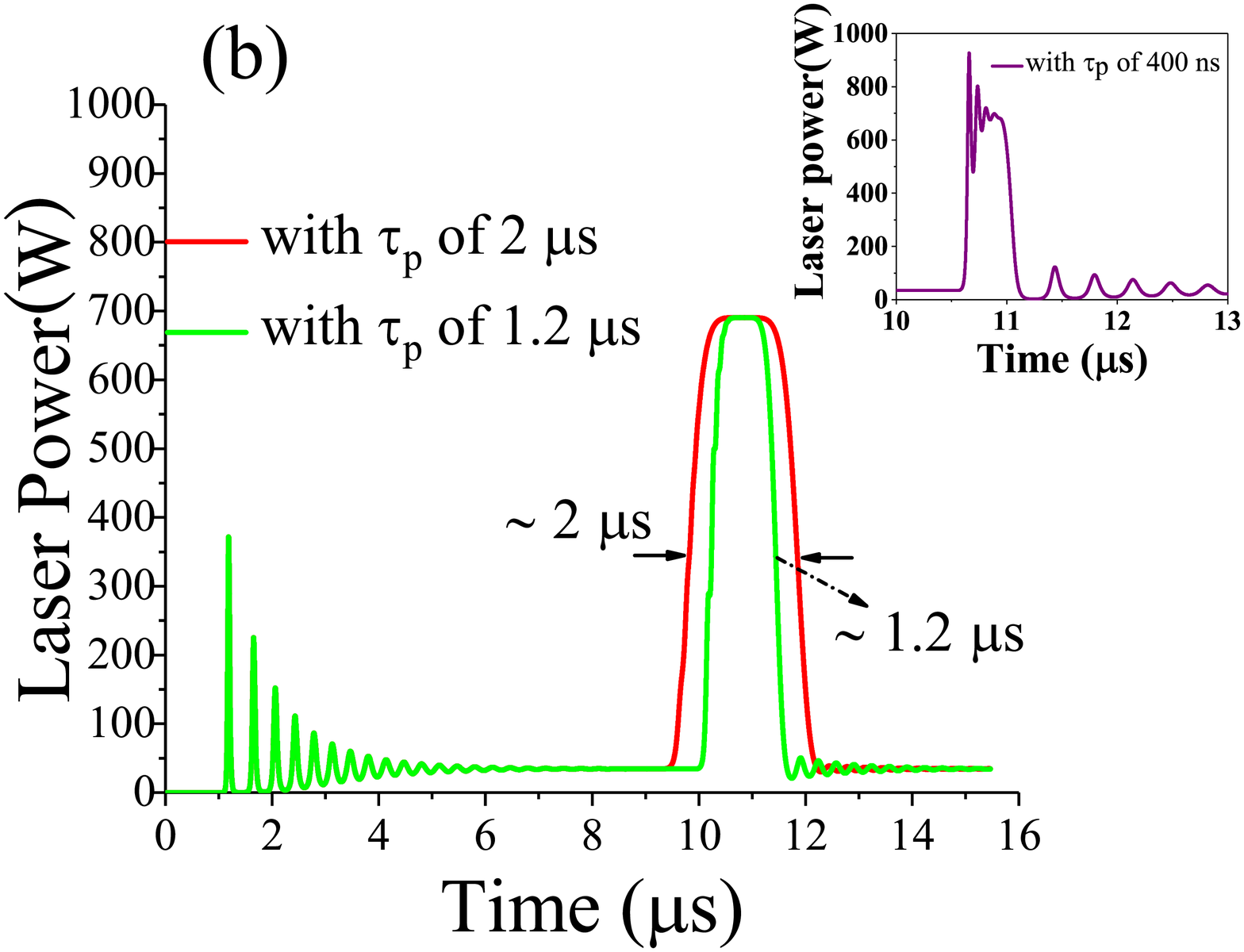}}
\centerline{ }
\end{minipage}
\vfill
\begin{minipage}{0.5\linewidth}
\centerline{\includegraphics[width=7cm]{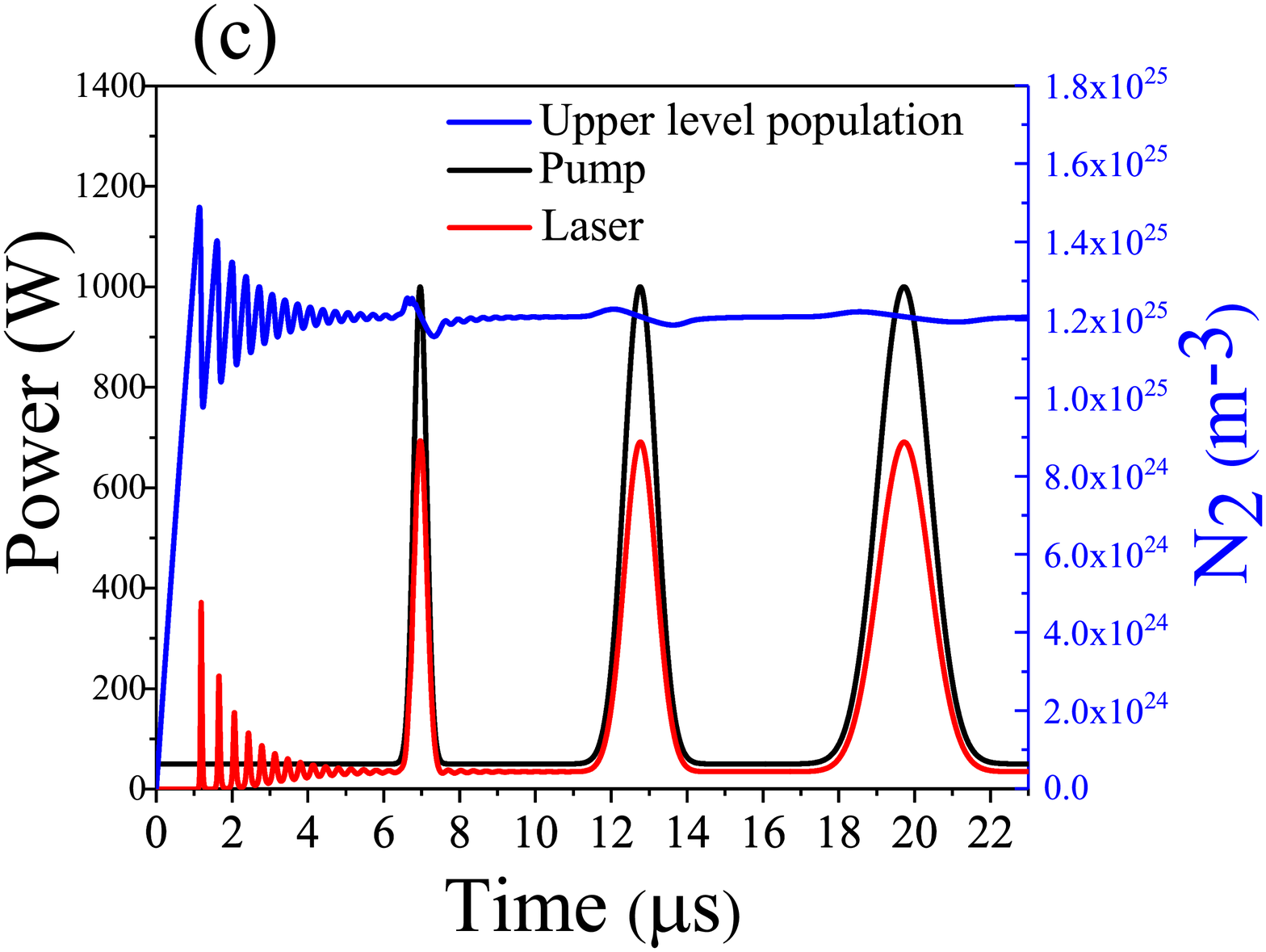}}
\centerline{}
\end{minipage}
\hspace{-0.35in}
\begin{minipage}{0.5\linewidth}
\centerline{\includegraphics[width=7cm]{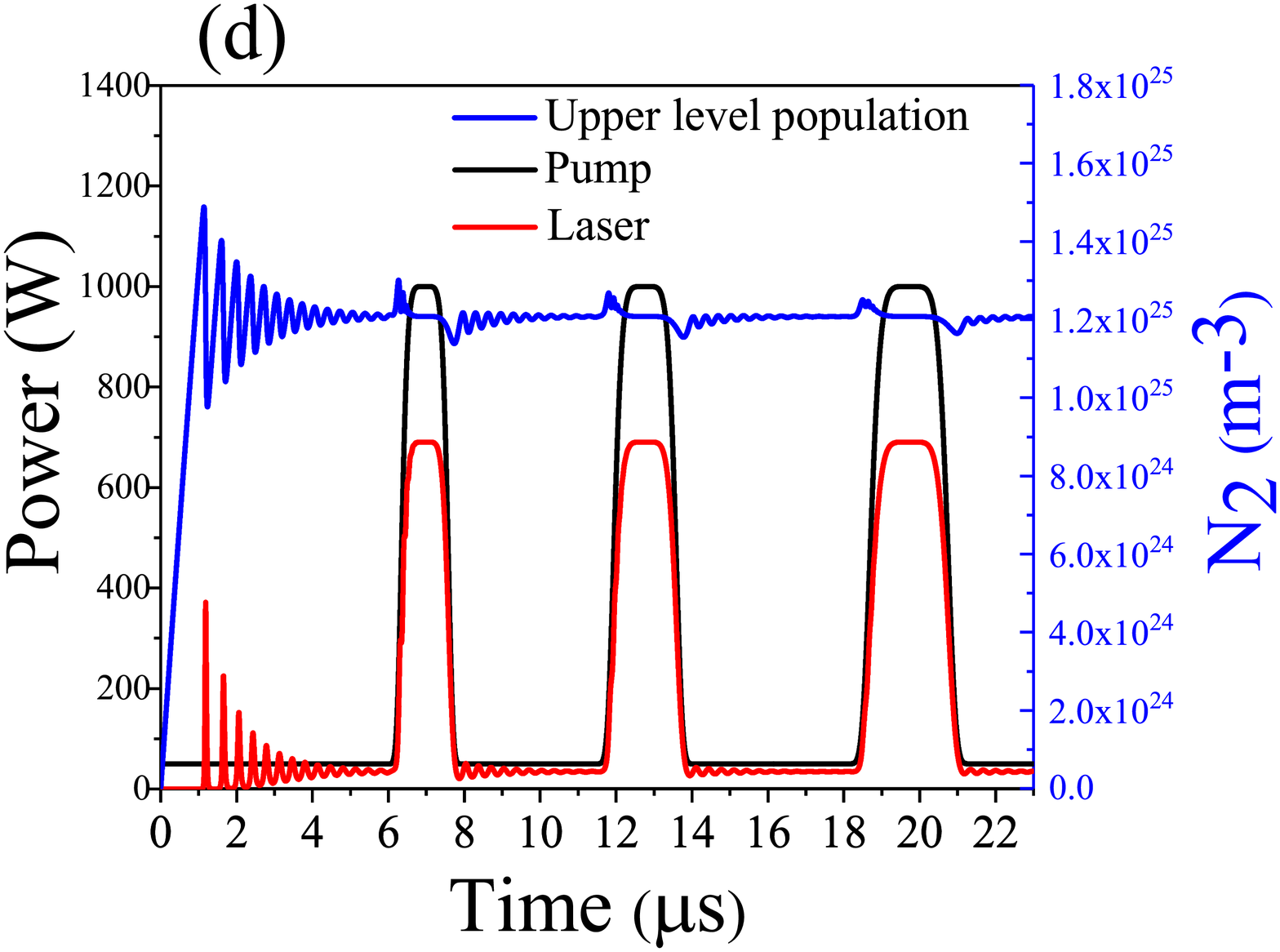}}
\centerline{ }
\end{minipage}
\caption{With 5 $\%$ pump power bias, temporal characteristics of output signal pulses when the pump pulses have a peak power of 1000 W with different durations and (a) Gaussion profiles, (b) 6-order super-Gaussian profiles. With 5 $\%$ pump power bias, temporal characteristics of the laser pulse trains and N$_2$(z=0,t) (at the position of z=0) when the pump pulse trains have (c) Gaussion profiles, (d) 6-order super-Gaussian profiles, with different durations. The duration of pump pulse is represented by $\tau _p$.}
\label{ss}
\end{figure*}

Instead of seeding signal laser at one side of fiber continuously, the seeded signal laser can be more easily obtained by switching on the bias pump firstly. It is just a CW pumped fiber laser with only bias pump operation. After experiencing relaxation oscillation the output laser reaches a CW steady state under bias-pumped power of 50 W, as seen in Fig. \ref{ss}(a). And then the signal laser in CW state serves as a seed for the signal pulse to be generated and pulsed pump with a peak power of 1000 W and Gaussion profile is switched on. The simulation results are shown in Fig. \ref{ss}(a). Stable signal pulses are produced and their durations are almost the same as those of the pump pulses depicted in Fig. \ref{ss}(a). Even if the duration of Gaussion pump pulse increases to 2 $\mu$s, stable pulse can also be realized without relaxation spikes appearing. 

When the profile of pump pulse is changed from Gaussion profile to 6-order super-Gaussian profile, the simulation results are shown in Fig. \ref{ss}(b) with other parameters the same as those of Fig. \ref{ss}(a). In the 6-order super-Gaussian profile case, spikes in the front edge of the output pulse cannot be eliminated completely with the pulsed pump duration of 0.4 $\mu$s. 
When the duration of 6-order super-Gaussion profile is broader than 1.2 $\mu$s, the chaotic spikes disappear and stable output pulses are achieved. Both shorter pulse duration and higher order super-Gaussion profile mean rapider power changing at the edge of pump pulse, which tends to result in relaxation oscillation.   

\begin{figure*}[htbp]
\centerline{\includegraphics[width=4in]{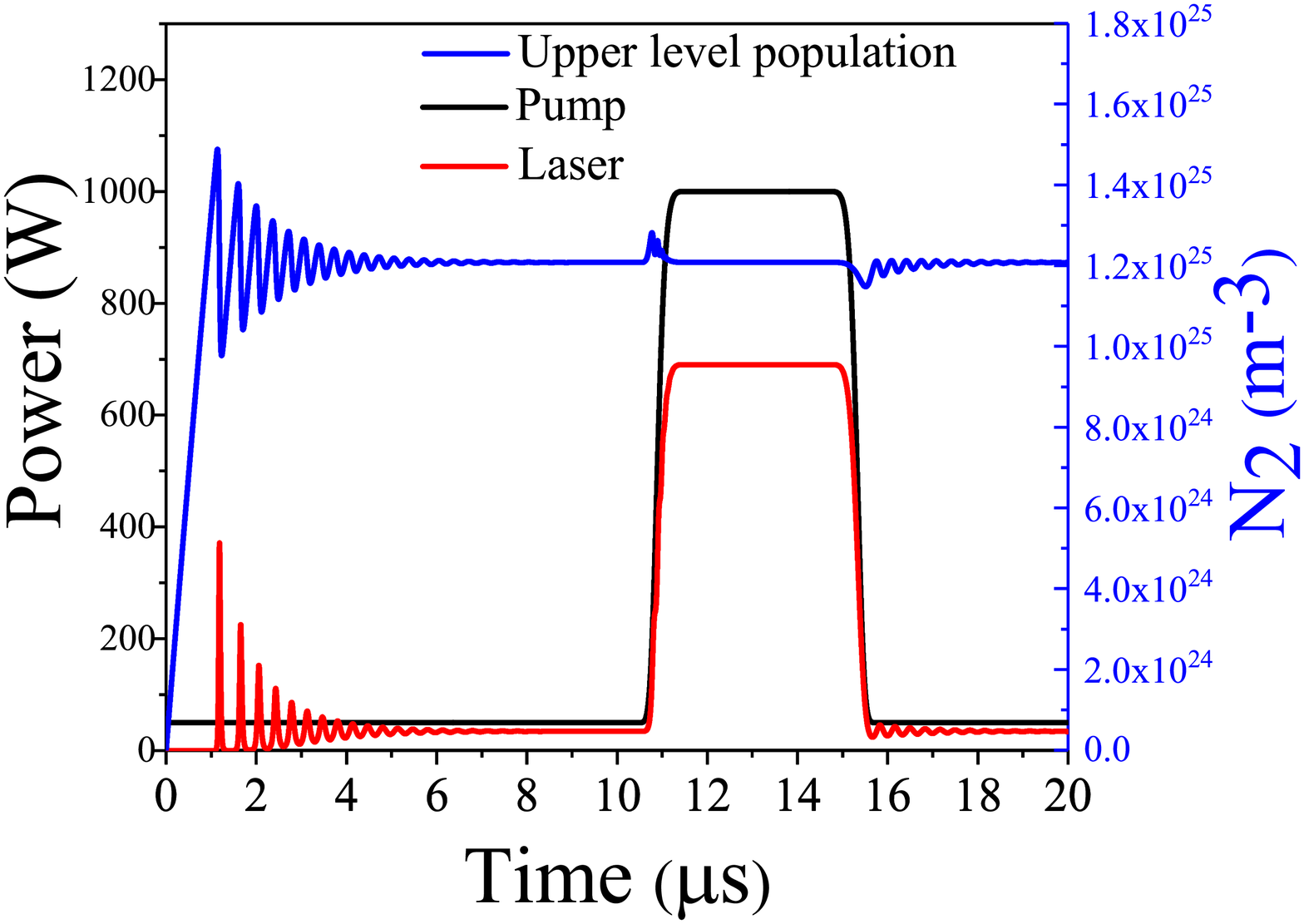}}
\caption{With 5 $\%$ pump power bias, temporal characteristics of the output pulse and N$_2$(z=0,t) (at the position of z=0) when the profile of pump pulse is flat-top with smooth rising and falling edges.}\label{sr0}
\end{figure*}

The simulation results in Fig. \ref{ss}(a) and (b) clearly show the output pulses can keep the same durations as those of pump pulses. Consequently, the output pulse duration can be tuned by managing the pump pulse duration. Stable output pulse trains with different durations are produced under a sequence of pump pulses with different durations, and the results are shown in Fig. \ref{ss}(c) with Gaussion pump profile and Fig. \ref{ss}(d) with 6-order super-Gaussion profile. Between two adjacent output pulses, the emission laser is nearly in a CW state with nonzero power, which severs as a seed for the next pulse to suppress the chaotic relaxation spikes. In addition, the profiles of output pulses are almost identical with those of pump pulses adopting 5 $\%$ bias pumping, as shown in Fig. \ref{ss}. It means that the profile of output pulse can be shaped by controlling the pump pulse profile. Therefore, we get a method to control the temporal shape of signal pulse in fiber laser. For example, laser pulse with a flat-top profile can be obtained by adopting bias-pumping technique and modulating the pump profile as a flat-top shape with smooth rising and falling edges, as illustrated in Fig. \ref{sr0}. In practice, the time-varying of pump power is easily modulated as a flat-top profile with smooth edges which can be approximated with a super-Gaussian profile.  

The temporal evolution of upper level population density N$_2$(z=0,t) at the position of z=0 is also simulated. With seeded laser suppressing the relaxation spikes, there are no population oscillation across the output pulses as shown in Fig. \ref{ss}(c), Fig. \ref{ss}(d) and Fig. \ref{sr0}. But slight population oscillation occurs at both edges of output pulses when the pump powers are changing fast, as seen the blue lines in Fig. \ref{ss}(d) and Fig. \ref{sr0}. The temporal characteristics of N$_2$(z=0,t) further indicate the mechanism of pulse formation in this novel pulsed fiber laser is not based on relaxation oscillation. The peak power of flat-top pulse in Fig. \ref{sr0} is lasting about 3 $\mu$s, which is actually in a CW steady state. Synchronization of temporal evolution between the output signal pulse and the pump pulse is realized with signal laser seeding, which is the mechanism of pulse-shaping in this novel pulsed fiber laser.    

Based on the new pulsing mechanism, bias-pumped gain-switched fiber laser has potential for generating stable pulses with tunable durations and controllable pulse shapes. Owing to the temporal evolution of output pulse is synchronous with that of pump pulse, the duration and temporal profile of output pulse are almost the same as those of pump pulse. Since no relaxation spikes occur under long pump duration, we donot have to rely on fast gain-switching technique to switch off/on the pump rapidly. Therefore, the control unit which monitors the laser output and control the pumping diodes in conventional gain-switched fiber laser can be left out in this novel pulsed fiber laser.

\section{Conclusion}

We consider the bias-pumped gain-switched fiber laser a novel pulsed fiber laser based on a new pulsing mechanism. The key factor for this novel pulsed fiber laser is the synchronization of temporal evolution between the output signal laser and the pump, which requires signal laser seeding with a certain power. Bias-pumping technique proposed by us is a convenient way to supply seed laser which can suppress relaxation oscillation. Stable output pulses with tunable durations can be obtained under CW bias pump combined with pulsed pump. The CW bias pump power is responsible for suppliying seed laser and pulsed pump is responsible for shaping the output pulse. With a certain signal power seeding, the output pulse can keep almost the same duration and temporal profile as those of the pump pulse, as long as the pump profile is smooth. Thus, the temporal characteristics of output pulse including duration and pulse shape can be controlled by management of the pump pulse. In addition, the control unit used to realize fast gain-switching in conventional gain-switched fiber laser can be left out in this novel pulsed fiber laser, which greatly simplifies the pulsed laser design. Based on the new pulsing mechanism and excellent output pulse characteristics, this kind of novel pulsed fiber laser may open up novel industrial and scientific applications.

\end{document}